# Trapping of light in solitonic cavities and its role in the supercontinuum generation

R. Driben,[1,2,*] A.V. Yulin,[3] A. Efimov,[4] and B.A Malomed[2]

[1]*Department of Physics & CeOPP, University of Paderborn, Warburger Str. 100, D-33098 Paderborn, Germany*
[2]*Department of Physical Electronics, Faculty of Engineering, Tel Aviv University, Tel Aviv 69978, Israel*
[3]*Centro de Física Teórica e Computacional, Universidade de Lisboa, Ave. G. Pinto 2, Lisboa 1649-003, Portugal*
[4]*Center for Integrated Nanotechnologies, Materials Physics and Applications, Los Alamos National Laboratory, Los Alamos, NM 87544, USA*
[*]*driben@post.tau.ac.il*

**Abstract:** We demonstrate that the fission of higher-order N-solitons with a subsequent ejection of fundamental quasi-solitons creates solitonic cavities, formed by a pair of solitons with dispersive light trapped between them. As a result of multiple reflections of the trapped light from the bounding solitons which act as mirrors, they bend their trajectories and collide. In the spectral-domain, the two solitons receive blue and red wavelength shifts, respectively. The spectrum of the bouncing trapped light alters as well. This phenomenon strongly affects spectral characteristics of the generated supercontinuum. Studies of the system's parameters, which are responsible for the creation of the cavities, reveal possibilities of predicting and controlling soliton-soliton collisions induced by multiple reflections of the trapped light.

**1. Introduction**

Various aspects of the dynamics of ultrashort pulses in photonic crystal fibers (PCFs), such as higher-order soliton fission and interaction of solitons with radiation, is subject of profound interest for fundamental studies and technological applications in photonics. The fission of higher-order solitons [1-3] is a key mechanism for the generation of ultrashort frequency-tuned fundamental solitons [4] and ultra-broadband optical supercontinuum (SC) [3, 5-10]. Recently, a model based on an optical-soliton counterpart of the Newton's cradle (NC) was introduced for understanding the mechanism of *N*-soliton fission under the action of the higher-order dispersion [11]. The mechanism, which remains relevant in the presence of the Raman and self-steepening effects, explains the discrete nature of ejections of fundamental quasi-solitons from the parental *N*-soliton temporal slot. It was also demonstrated that the delay between the ejections of individual solitons can be controlled by means of the strength of the third-order dispersion. Along with ejected solitons *N*-soliton fission gives rise to emission of strong dispersive radiation. Quasi-solitons with higher peak powers are ejected before their weaker counterparts, and they experience stronger Raman self-frequency shifts [12]. Due to the interaction with the radiation, weaker solitons can acquire an additional acceleration, which leads to their collisions with the stronger solitons ejected earlier [13]. As these interactions strongly influence the spectral broadening of the pulse at all the stages of the SC formation, a challenging objective is to develop reliable control over these phenomena. Among particular types of the interactions involving solitons and radiation, notable are various forms of the four-wave mixing (FWM) [14-18], trapping of dispersive waves by a soliton [19], formation of bound states of solitons [20-22], soliton fusion [23], etc. The interactions between dispersive waves and solitons were also studied in the context of all-optical switching and rogue wave formation [24, 25]. In a recent work [26], dispersive waves trapped in an effective especially predesigned cavity, created by two separate solitons acting like mirrors, were considered, and it was shown that the FWM of the solitons with the dispersive waves strongly affects the dynamics of both the solitons and the trapped waves. In particular, this interaction may result in broadening or narrowing of the spectrum of the trapped waves [26, 27]. Furthermore, the radiation-induced attraction ("Casimir force") between the solitons eventually causes them to collide.

The aim of the present work is to investigate how the FWM between the solitons and dispersive waves in two-soliton cavities, which emerge following the fission of *N*-solitons, affects the ensuing supercontinuum generation. It will be demonstrated how all the components comprising the cavity and experiencing frequency modifications affect the overall form of the emerging SC from the very onset of its generation.

## 2. Light trapping by solitonic cavity after the fission of higher order soliton.

For the sake of clarity, we disregard the Raman self-frequency shift (which is a valid assumption for a hollow-core PCFs filled with Raman-inactive Xenon gas [28]) and dispersion terms above the third order. In general, the cavity dynamics reported below remains valid if those effects are taken into regard, as shown in the last part of the paper.

Under the conditions outlined the dynamics of an ultrashort pulse's amplitude A is governed by the generalized nonlinear Schrödinger equation (NLSE) [7, 28], which includes the second- and third-order dispersion (TOD) terms with respective coefficients $\beta_2$ and $\beta_3$, and the cubic nonlinear term, along with its self-steepening part:

$$\frac{\partial A}{\partial z} = -\frac{i\beta_2}{2}\frac{\partial^2 A}{\partial T^2} + \frac{\beta_3}{6}\frac{\partial^3 A}{\partial T^3} + i\gamma\left[A|A|^2 + \frac{i}{\omega_0}\frac{\partial}{\partial T}\left(A|A|^2\right)\right] \qquad (1)$$

With the retarded time $T = t - z/v_g$ and the the group velocity $v_g$ of the light at the carrier frequency $\omega_0$. Fig. 1 demonstrates the fission of a *N*-soliton with $N = 12$ and 800-nm central wavelength, as produced by a numerical simulation of Eq. (1). The input is taken, accordingly, as $u_0 = \sqrt{P_{in}}\operatorname{sech}(T/T_0)$ with $T_0 = 50$fs ($T_{FWHM} \sim 90$ fs) and $P_{in} = N^2 P_0$, where $P_0 = 56$ kW. The fiber parameters are $\beta_2 = -0.0021$ ps$^2$/m, $\beta_3 = 5.235 \cdot 10^{-6}$ ps$^3$/m and $\gamma = 1.5 \cdot 10^{-5}$ W$^{-1}$m$^{-1}$.

After the breakup of the injected *N*-soliton, a strong Cherenkov [8] dispersive wave packet is emitted at $Z = 0.1$ m, which appears at the blue edge of the total spectrum around 450 nm and is not involved in the process described below (see Fig. 1). It is the light from the periodic chain of pulses [11] propagating in a bound state, between the ejections of fundamental quasi-solitons, that gets partially trapped between the solitons, bouncing between them, starting from $Z = 0.16$ m [Fig. 1 (a)]. Although the peak intensities of the trapped wave packets are small in comparison to those of the solitons (5-6 times lower), they play a major role in the subsequent dynamics of the two solitons and, consequently, in the spectral evolution of the generated SC [Fig. 1(b)]. Multiple quasi-elastic [29] collisions of the trapped dispersive wave with the solitons cause acceleration and deceleration of the solitons, manifested in bending of their trajectories, [Fig. 1(a)] i.e., effective tilting of the mirrors in the respective cavity picture. In the spectral domain, the curvature of the trajectories is represented by a blue shift for the first soliton (from 1010 nm to 930nm), and a red shift for the second soliton (from 920 nm to 1030nm), as spectral filtering of individual solitons reveals [the dotted black and solid magenta curves in Fig. 1(b) demonstrate the evolution of the central wavelengths of the solitons]. The spectral region between 460 nm and 620 nm, where the reflections of the trapped waves occur, is shown by a dashed white rectangle in Fig. 1(b), and it is discussed in more detail below.

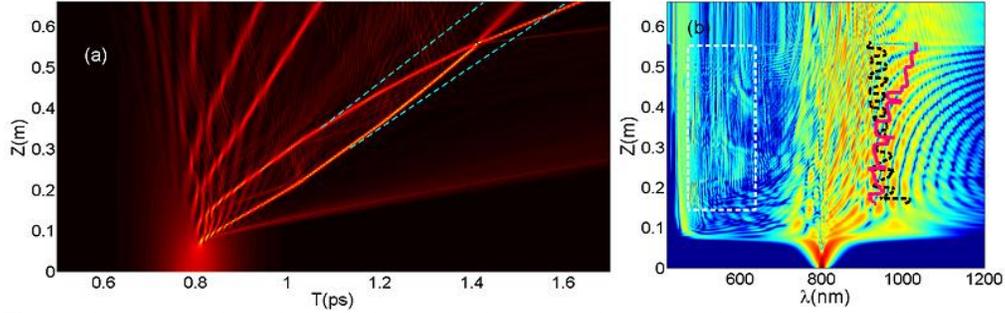

**Fig. 1.** *Solitonic cavity obtained after the fission of a 12- soliton with $T_{FWHM}$ =90 fs and peak power of 8MW, in the PCF filled with a Raman-inactive gas: (a) temporal- and (b) spectral-domain representations of the pulse's dynamics. Dashed blue lines in (a) designate trajectories of the two solitons produced by artificially filtering out all the radiation trapped between them. Dashed black and solid magenta lines in (b) designate the evolution of spectral maxima of the first and the second solitons, respectively, starting from Z=0.16m. The spectral region of multiple bouncing of dispersive waves is highlighted by a dashed rectangle.*

To prove the critical role of the interaction with the dispersive waves in bending of solitons' trajectories and their spectral shifts, an additional simulation was performed, in which we filter out all the radiation between the two solitons at $Z = 0.16$ m. In this case, the solitons continued their propagation with undistorted trajectories, as shown in Fig. 1(a) by the dotted blue lines. As a result of the induced acceleration of the second soliton and deceleration of the first one, the two collide at $z = 0.56$ m. At some distance before the collision, the "soliton mirrors" degrade, allowing the trapped radiation to escape. The narrowing of the effective waveguide created by the two solitons strongly resembles the effect of the tapered waveguides. The nature of the interaction between the dispersive waves and solitons dealt with here is quite different from [13] where the dispersive waves emitted by the first soliton collided with the second one and mostly passed through it without being reflected back to the first soliton and affecting its trajectory or central wavelength.

To further clarify the present effect, we have performed another numerical experiment, by filtering out everything from $Z = 0.16$, except the cavity consisting of the two solitons with the trapped light between them, see Fig. 2(a). Fig. 2(b) shows the evolution of the trapped-light spectrum while the spectral lines of the solitons are located far away, between 900 and 1000nm, as seen in Fig. 1(b). It can be observed from both panels of Fig. 2 that the trapped light experiences strong spectral changes as it collides with the solitons, similar to the case of a solitonic cavity specially constructed in [26]. The condition of the resonant scattering [15] for the case referred to as phase-insensitive in [18] can be written in the form:

$$\frac{\beta_2}{2}\delta_i^2 + \frac{\beta_3}{6}\delta_i^3 + \left(\beta_2\delta_s + \frac{\beta_3}{2}\delta_s^2\right)(\delta - \delta_i) = \frac{\beta_2}{2}\delta^2 + \frac{\beta_3}{6}\delta^3 \qquad (2)$$

where the frequency deviations from the one corresponding to the dispersion curve reference wavelength-$\lambda_0$ are: $\delta_i = 2\pi c(\frac{1}{\lambda_i} - \frac{1}{\lambda_0})$, $\delta_s = 2\pi c\left(\frac{1}{\lambda_s} - \frac{1}{\lambda_0}\right)$, $\delta = 2\pi c\left(\frac{1}{\lambda} - \frac{1}{\lambda_0}\right)$, $\lambda_i$, $\lambda_s$ and $\lambda$ being the wavelengths of the incident wave, the soliton and the scattered wave, respectively. The left-hand side of Eq. (2) can be referred to as the dispersion characteristic of the FWM, while the right-hand side represents the dispersion of the fiber. Solutions of Eq. (2) give wavelengths at which the FWM terms are in resonance with the system's eigenmodes. For example, for the first reflection shown in Fig. 2 we deduce from Eq. (2) that the trapped light packet, centered around $\lambda_p$= 490nm, bouncing off the first soliton (centered around $\lambda_s$= 950 nm ) is reflected into a spectral region around $\lambda$= 596 nm. The fact that the solitons have slightly different wavelengths which are changing in the course of the propagation explains the observation that the cascaded re-scattering of the trapped waves is not exactly periodic, leading to the modification of the trapped radiation.

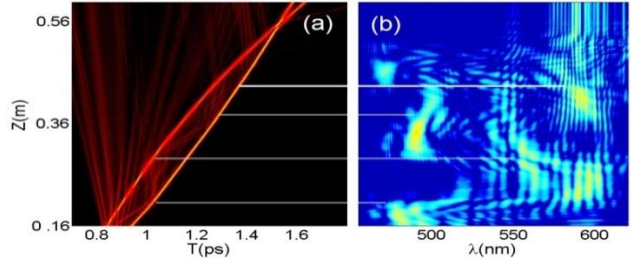

**Fig. 2**. *Dynamics of the cavity filtered at Z=0.16m, so that only the two solitons and the dispersive waves trapped between them are kept in the system, in the temporal (a) and spectral (b) domains. Four horizontal lines relate reflections of the trapped light from the solitons with the reciprocal changes in the spectrum of the trapped light.*

To observe the action of the soliton cavity, one has to carefully choose system parameters. For very small values of the TOD parameter $\beta_3$, the fission leads to a decomposition of the $N$-soliton into $N$ distinct fundamental solitons without trapping any light between them, as predicted for the (nearly) integrable NLSE [1,2]. In this case, the peak power of each fundamental soliton emerging after the splitting of the N-soliton is given by $P_j = P_0 (2N - 2j + 1)^2$ [1,2]. On the other hand, under larger values of $\beta_3$, the fission is dominated by the NC mechanism [11], with solitons ejected in a step-like sequence, see Fig. 3(a-c). However, if the relative TOD is too strong, ejections of the fundamental quasi-solitons occur with a very large relative delay along $Z$ [11], leaving little opportunity for the light trapped between the ejected solitons to mediate the attractive interaction.

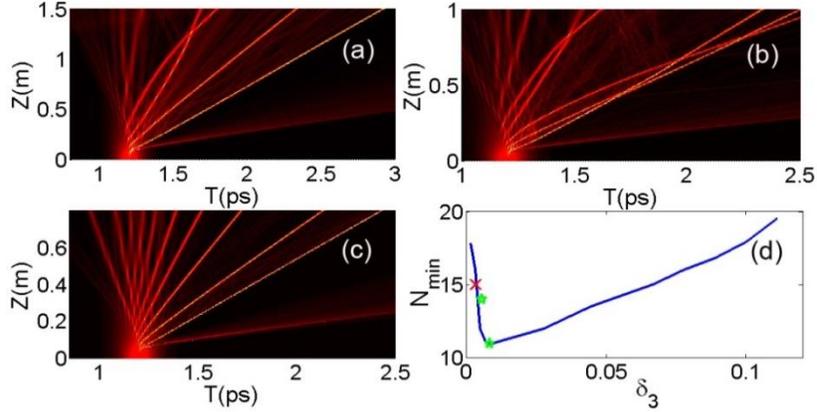

**Fig. 3**. *(a-c) N-soliton fission in the PCF filled with a Raman-inactive gas. (a) Fission of a 11-soliton with $T_{FWHM}$ =90 fs and $\beta_3$ =5.235·10$^{-6}$ ps$^3$/m; (b) fission of a 14-soliton with $T_{FWHM}$ =90 fs and $\beta_3$ =3.49·10$^{-6}$ ps$^3$/m; (c) fission of a 15-soliton with $T_{FWHM}$ =90 fs and $\beta_3$ =2.1·10$^{-6}$ ps$^3$/m. (d) Smallest order N necessary to build a solitonic cavity, as a function of the relative TOD strength,-$\delta_3$.*

The solitonic cavity can be formed by not only the first two ejected solitons, but by another pair as well. Fig. 3(a) demonstrates that, despite evident multiple reflections of the trapped light between the first two solitons, no significant bending of their trajectories is observed for this pair. This is due to a strong initial frequency difference received by these two solitons, while they were being ejected. However, the interaction between the third and fourth solitons via the light trapped between them overtakes their initial spectral separation and the two solitons get attracted to each other. The effective cavities may involve additional solitons as Fig. 3(b) reveals.

For each normalized value of the relative TOD strength, defined as $\delta_3 = \beta_3/(6|\beta_2|T_0)$, there exists a minimum value, $N_{min}$, of the input-soliton's order $N$ for which the light resonating between any ejected soliton pair induce effective attraction and subsequent collision between them. At $N < N_{min}$, no collisions occur, e.g. see Fig. 3(c), since for smaller N there will be less fundamental solitons ejected and the separation between the ejected solitons in space and in frequencies will be large enough to overpower the attraction mechanism driven by the bouncing light in between. $N_{min}$ is plotted versus $\delta_3$ in Fig. 3(d) in the range of $0.0016 < \delta_3 < 0.11$. The graph shows that the absolute minimum of the input soliton order is $N \approx 11$, for which the solitons collide if the fission takes place at $\delta_3 = 0.0083$. The two green marks near the left segment of the curve pertain to the cases shown in Fig. 3(a,b), while the red mark pertains to case (c), when no collision occurs.

Now we discuss the role of the higher-order nonlinear and dispersion terms in the generalized NLSE. The shock term, which has been included in Eq. (1), facilitates the ejection of quasi-solitons from the NC, therefore it facilitates the attraction and collisions between the solitons. If the shock term is dropped, the first "collapsing cavity" will appear with minimal $N = 16$ instead of $N = 12$ in Fig. 1, if the same values of other parameters are used. In the presence of the standard oscillatory Raman term in the extended NLSE which describes regular silica PCFs [8,10], red-shifted accelerated solitons tend to separate from each other faster. Nevertheless, the mechanism of the soliton-soliton attraction through the trapped light can overpower this tendency, as seen in Fig. 4(a). Further, the effect reported here is not specific solely for the TOD, and may be observed as well in the presence of higher-order dispersion terms. In the case when the dispersion terms up to the seventh order are included (taken from [10]), the effect looks qualitatively the same, see Fig. 4(b,c). Also for fibers with the leading TOD term, the dependence of $N_{min}$ on the TOD strength will look qualitatively similar to the one presented in Fig. 3(d) unaffected by higher order dispersion terms.

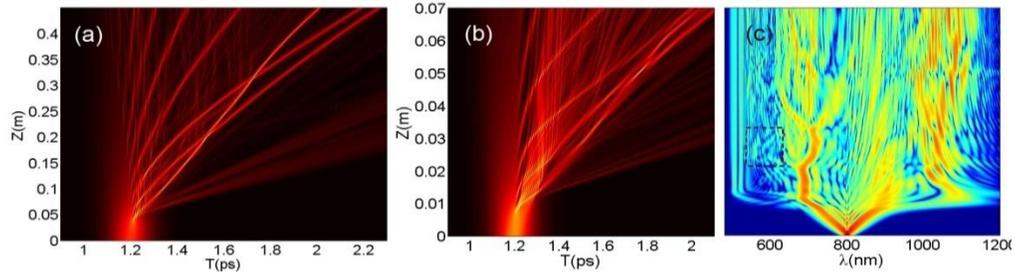

**Fig. 4** (a) Fission of a 18- soliton with $T_{FWHM}$ =90 fs and $\beta_3$ =5.235·10$^{-6}$ ps$^3$/m, in the regular silica PCF with the standard Raman term added to Eq.(1). (b) Temporal and (c) spectral SC evolution resulting from the fission of a 50-soliton with $T_{FWHM}$ =90 fs in a regular silica PCF, with the Raman term and dispersive ones up to seventh order included in Eq.(1). The spectral region of two consecutive bouncing of the trapped light from the two first solitons is highlighted by the black rectangle.

## 3. Conclusions

We have demonstrated the effect of the cavity created by two solitons and light trapped between them, which forms after the fission of the initial *N*-soliton. As a result of multiple reflections of the dispersive waves from the bounding solitons, which act as mirrors in a cavity, the solitons experience strong bending of their trajectories and eventually collide. In the spectral domain, this manifests itself in blue and red wavelength shifts of the two solitons. The spectrum of resonating dispersion waves also changes as they bounce each time from the solitons with varying wavelengths. The phenomenon strongly influences spectral characteristics of the generated supercontinuum. The systematic study has identified the minimum order *N* of the input solitons, above which the mutual attraction and ensuing collision of the solitons occur. The mechanism, which was first demonstrated in the Raman-free medium, remains valid in a model of a regular glass fiber, with all higher-order linear and nonlinear terms included.